\documentclass{revtex4}
\begin{document}

\title{Exponential energy growth due to slow parameter oscillations in quantum mechanical systems}
\author{Dmitry Turaev}
\email{dturaev@imperial.ac.uk}
\affiliation{Imperial College, London, SW7 2AZ, UK\\Lobachevsky University of Nizhny Novgorod\\Joseph Meyerhoff Visiting Professor (Weizmann Institute of Science)}
\begin{abstract}
It is shown that a periodic emergence and destruction of an additional quantum number 
leads to an exponential growth of energy of a quantum mechanical system subjected to a
slow periodic variation of parameters. The main example is given by systems (e.g., quantum billiards and quantum graphs) with
periodically divided configuration space. In special cases, the process can also lead to a long period of cooling that precedes the
acceleration, and to the desertion of the states with a particular value of the quantum number.
\end{abstract}
\maketitle
It is well known \cite{Hu,BF,LL} that the state of a quantum mechanical system 
subjected to a slow variation of parameters changes adiabatically: for a system in a given energy eigenstate
the transition amplitude to other energy levels due to a slow change of parameters is small. More precisely, if the dimensions 
are scaled such that the gap between the neighboring energy levels is of order $1$ and the parameters of the system
change periodically with the speed of order $\varepsilon$, then the system in a state with definite instantaneous energy 
will, after each period, return to the $O(\varepsilon)$-vicinity of the initial state (with a possible phase shift \cite{Be}).
This continues for at least $O(\varepsilon^{-1})$ periods, with probability close to 1. In this paper, we describe a mechanism of the energy levels
crossing, for which the system's response to the slow variation of parameters is still adiabatic (i.e.,
starting with a definite energy state, the system, with probability close to 1, closely follows a state
of definite instantaneous energy for a long time); however, the new state after each period of the parameters oscillations
is, typically, different from the initial one, and the averaged energy gain per period is positive.

Level crossing is usually associated with symmetries in the system, e.g., with the integrability of the corresponding classical system \cite{NW,BT,LL},
but our construction is different. It is based on a periodic emergence and destruction of an additional quantum number in the slowly
perturbed system. This can be achieved in a variety of ways, e.g. by imposing a magnetic field whose
spatial coherence properties depend on time, like in Eq.~(\ref{mf}). Our basic example is a system with
a periodically disconnected configuration space. It may be a quantum billiard - a free particle confined
to a bounded domain $D\subset R^d$ (see Refs. \cite{BT,BR,BaR,LLP,Ze} for references on quantum billiards and \cite{GS,qg} on quantum graphs). If $d\geq 2$,
the domain $D$ can be slowly deformed in such a way that at some moment two boundary arcs touch,
after which $D$ is divided into two parts, $D_1$ and $D_2$ \cite{GRST11}. The two domains then evolve separately until
they reconnect again, and the process repeats periodically. In general, there is no symmetry between $D_1$ and $D_2$,
so at the separation moment the energy eigenstates are divided into two groups. The eigenstates from group I are
eigenfunctions of the Laplacian in $D_1$ and identically zero in $D_2$; the eigenstates from group II
are eigenfunctions of the Laplacian in $D_2$ and zero in $D_1$, cf. \cite{Hu}. With time, the shape
of $D_{1,2}$ changes and, typically, there is no level crossing within each group, but
the levels from different groups can cross. As we show below, this can lead the system at the moment of reconnection
to the energy level different from the initial one. Moreover, the corresponding energy values recorded at the beginning of each
period grow, on average, exponentially with time.

One extends the class of examples with the divided configuration space by adding a (possibly time-dependent)
potential inside the domain $D$. The same scheme is also applicable in the 1-dimensional case: one can consider
a time-dependent quantum graph whose connectivity changes adiabatically with time. This can be achieved by cutting
some edges in an adiabatic manner (by introducing the semi-penetration boundary conditions like in
Eq.~(\ref{bc1}) below). If the graph is periodically divided into disconnected parts and reconnected again,
then one should, in general, expect the particle in such graph to experience an exponential acceleration.

The exponential Fermi acceleration in classical billiards with periodically divided configuration space
was discovered in Ref. \cite{GRST11}. It was shown in Refs. \cite{Mush,Ba,PT1,PT2} that this is a partial case of a general phenomenon:
a slow periodic variation of parameters of a non-ergodic Hamiltonian (classical) system of an arbitrary nature
leads, generically, to an exponential growth of energy. A quantum mechanical analogue of this principle
would be that if at high enough energies
the gaps between energy levels become all small of order $\varepsilon$ (in order to ensure non-vanishing amplitudes
of the transition between the levels), then the quantum behavior should, probably, mimic the classical one:
a fast energy growth should, typically, be expected if the classical limit system is non-ergodic.
We do not know to which extent this conjecture is true. Moreover, there exist systems 
for which the spectral gaps do not uniformly tend to zero as the energy grows (this happens e.g. for quantum 2-dimensional billiards, 
as follows from the Weyl law \cite{We}). The quantum acceleration
construction presented in this paper does not rely on the classical dynamics properties in the high energy limit
and can ensure the exponential energy growth starting even with the lowest energy state.

We begin with a toy model of a particle confined in a segment of a straight line; this is the simplest case of
both a quantum billiard and a quantum graph. Consider a particle in a segment $[-1,a]$ whose end point $a$ moves
slowly and  periodically in time, and for a portion of the period the segment is divided into two parts, so the
particle cannot penetrate from one part to another. This is described by the Schr\"odinger equation
$$i\psi_t = -\psi_{xx}+V(x) \psi,$$
with the boundary conditions $\psi(-1)=\psi(a)=0$ where $a(\varepsilon t)>0$ is a $T$-periodic function
and $\varepsilon>0$ is sufficiently small. We introduce a barrier at $x=0$
by allowing a discontinuity of the first derivative at zero and introducing the following
time-dependent boundary condition:
\begin{equation}\label{bc1}
\alpha(\varepsilon t) \psi(0) + (1-\alpha(\varepsilon t))(\psi_x(+0)-\psi_x(-0))=0.
\end{equation}
At $\alpha=0$ the barrier is absent, while at $\alpha=1$ the segments $x<0$ and $x>0$ are completely separated.
It is easy to check that the operator $\psi \to \psi_{xx}$ with this boundary condition remains self-adjoint
at each moment of time. We do not discuss a possible physical meaning behind this separation mechanism;
our goal is just to be sure that the adiabatic separation of a segment into two disjoint parts is
mathematically feasible.

We assume that there is a total separation for a slow-time interval $[\tau_1,\tau_2]\subset (0,T)$, i.e.
$\alpha\equiv 1$ for $\tau:=\varepsilon t\in[\tau_1,\tau_2]$. We choose $V(x)=-1$ at $x<0$ and $V(x)=0$ at $x>0$
(this just helps us to make the computations explicit). During the separation interval, at each moment of $\tau$
there are two groups of energy eigenfunctions: the ``left'' states $\psi_n^-(\tau)$ ($n\geq 1$)
concentrated completely at $x\in (-1,0)$ and the ``right'' states $\psi_m^+(\tau)$ ($m\geq 1$)
concentrated at $x\in(0,a(\tau))$. The corresponding
energy levels are $E_n^-=-1+\pi^2n^2$ and $E_m^+=(\frac{\pi m}{a(\tau)})^2$.
During the slow motion of the boundary, the energy levels within each group do not intersect,
so we may assume that if the system is in an eigenstate $\psi_n^-(\tau)$ or $\psi_m^+(\tau)$ at $\tau=\tau_1$,
then it remains in this state at $\tau=\tau_2$, i.e., we neglect the order $\varepsilon$ amplitudes of the transition
between the eigenstates.

After the reconnection at $\tau=\tau_2$ the division into the two groups becomes
meaningless. We order the eigenstates by their energy; it is obvious that the state $\psi_n^-(\tau_2)$
acquires the number $k=n+\left[a(\tau_2)\sqrt{n^2-\frac{1}{\pi^2}}\right]$ in this total order (i.e.,
$k$ equals $n$ plus the number of the right states with the energies smaller than $E_n^-$), while the state
$\psi_m^+(\tau_2)$ acquires the number $k=m+\left[\sqrt{(m/a(\tau_2))^2+\frac{1}{\pi^2}}\right]$
(here the square brackets denote the integer part; note also that we assume a generic choice of $a(\tau)$, so
that $E_n^-\neq E_m^+$ for any $m$ and $n$ at the moments of the reconnection and separation).
Between the reconnection and the next separation at $\tau=\tau_1+T$ the system changes adiabatically, so
we may assume that the quantum number $k$ is conserved. At the separation moment the eigenstate with the
number $k$ is the right state with the number $m$ if $k=m+\left[\sqrt{(m/a(\tau_1))^2+\frac{1}{\pi^2}}\right]$ and
the left state with the number $n$ if $k=n+\left[a(\tau_1)\sqrt{n^2-\frac{1}{\pi^2}}\right]$. These formulas
take a particularly simple form if we, for example, choose $a(\tau)$ such that $a(\tau_1)=1$ and $a(\tau_2)=3$.
Then the left state $\psi_n^-$ has number $k=4n-1$ at the reconnection moment and number $k=2n-1$ at the separation moment,
while the right state $\psi_m^+$ has number $k=2m$ at the separation and the number
$k=m+[m/3]<2m$ at the reconnection.

Thus, the states with odd $k$ become left states after the separation and, after the reconnection, they
acquire a higher, and still odd, value of $k_{new}=2 k_{old}+1$. The states with even $k$ become right,
and acquire a strictly lower value of $k$. As we see, even though the system changes adiabatically all
the time, and absolutely no jumps between the energy levels is assumed, the level's number $k$ changes with each
period of the adiabatic oscillations. In our example, if we start with a right state, it will,
eventually, become left; we also have that the left states gain energy exponentially (the left level's numbers
double with each period). Therefore, if we start with a superposition of a finitely many eigenstates,
then after finitely many periods we will, with accuracy of order $\varepsilon$, have a superposition
of only left states, and the energy of the system will start growing exponentially with time.

Let us show that the exponential energy growth does not disappear if we change the details of the above
construction, and is a robust phenomenon for a general class of systems with a periodically divided
configuration space. Even more generally, we consider a quantum system with parameters which oscillate slowly enough
(so we assume that jumps between different energy levels do not occur), and
let for a part $[\tau_1,\tau_2]$ of the oscillation
period an additional quantum number emerge, so the energy eigenstates are divided into two groups, I and II,
such that the transition between group I and group II states is forbidden at $\tau\in [\tau_1,\tau_2]$. At each moment of
time we order the eigenstates by their energy $E_k$, $k=1,2,\dots$, and introduce the indicators $\sigma_j(k)$
as follows: $\sigma_j(k)=1$ if the eigenstate $\psi_k$ belongs to group I at $\tau=\tau_j$, and $\sigma_j(k)=-1$
if $\psi_k$ belongs to group II at $\tau=\tau_j$ ($j=1,2$). The two indicator sequences
$\sigma_{1,2}$ completely determine the energy evolution in the adiabatic approximation. Indeed,
at $\tau=\tau_j$, if the state $\psi_k$ is group I with the number $m$, then there are exactly $m$
group I and $(k-m)$ group II states with the energies not exceeding $E_k$, so $S_j(k):=\sigma_j(1)+\dots+\sigma_j(k)=2m-k$;
if $\psi_k$ is a group II state with the number $n$, then there are $n$
group I and $(k-n)$ group II states with the energies not exceeding $E_k$, so $S_j(k)=k-2n$. Thus, after the separation
moment $\tau=\tau_1$ the state $\psi_k$ becomes the group I (if $\sigma_1(k)=1$) or group II (if $\sigma_1(k)=-1$)
state with the number $n$ or $m$ equal to $(k+\sigma_1(k)S_1(k))/2$. After the reconnection at $\tau=\tau_2$
this state acquires the new number $\bar k$ such that
\begin{equation}\label{kbk}
\!\!\bar k+\sigma_2(\bar k) S_2(\bar k)=k+\sigma_1(k)S_1(k)\mbox{ and } \sigma_2(\bar k)=\sigma_1(k).
\end{equation}
By construction, this formula completely determines the change in the energy level after each period
of the parameter oscillations: the new level number $\bar k$ is a function of
the previous level number $k$, and vice versa. Thus, given an initial energy level number $k_0$,
the iteration of the rule (\ref{kbk}) provides a uniquely defined trajectory $k_s$ -
the sequence of the values of $k$ at the beginning of each period
(the same is true for backward iterations).

There are only two logically possible types of trajectories for the energy level number:
loops, when the system returns to the same energy level after a finite number of periods, and
unbounded trajectories, when the level number $k_s$
tends to infinity as the number of periods $s$ grows (in the latter case the level number will tend to infinity
also backwards in time). What happens for a given initial state of
a particular system, this depends on the detailed structure of the group I and II energies spectra
at the moments of separation and reconnection. This structure can be essentially arbitrary: while
the asymptotic behavior at large energies can be prescribed by the Weyl type formula, it is easy to
build, for any given $N$, a potential in any given domain such that the first $N$ energy eigenvalues
for a particle in this potential would take any given values, cf. \cite{ZM}. Thus, if the class of systems under consideration is sufficiently large,
we may think of the energy spectra of the group I and II states as random, i.e. the indicator sequences
$\sigma_1$ and $\sigma_2$ can be viewed as realizations of a certain random process.

The simplest model for $\sigma_1$ and $\sigma_2$ is given by sequences of independent random variables.
If $\beta$ is the probability of $\sigma_1(k)=1$, and $\gamma$ is the
probability of $\sigma_2(k)=1$, then $S_1(k)\sim (2\beta-1)k$ and $S_2(\bar k)\sim (2\gamma-1)\bar k$.
Now, Eq. (\ref{kbk}) gives $\bar k \sim \frac{\beta}{\gamma} k$ with probability $\beta$ (this corresponds to $\sigma_1(k)=1$),
and $\bar k \sim \frac{1-\beta}{1-\gamma} k$ with probability $1-\beta$ (corresponding to $\sigma_1(k)=-1$).
It follows that
$$\rho:=E(\ln\bar k - \ln k) \sim \beta \ln \frac{\beta}{\gamma} + (1-\beta) \ln \frac{1-\beta}{1-\gamma}.$$
If $\beta\neq\gamma$, then $\rho>0$ \cite{GRT}, i.e. the average energy gain per period is strictly positive. Therefore,
the probability to return to the initial value of $k$ falls exponentially with the number of periods.
This means that while several short loops may exist, long loops are rare, and a typical trajectory
of the energy level number is unbounded. By the law of large numbers, we have $\ln k_s \sim \rho s$ for a typical
realization of the process under consideration. It is natural to assume
$E_k\sim k^\nu$ for some $\nu>0$, thus the linear growth of $\ln k$ with time implies the exponential growth of the energy $E_k$. Obviously, the positivity of the gain
$\rho$ cannot be violated by small changes to the statistics of the transitions between the energy levels at the
separation and reconnection moments; it will persists even if we allow for a non-zero amplitude of transition
from one to several levels with a small spectral gap, provided such events are sufficiently rare. Thus, the exponential
energy growth should be a generic phenomenon in the adiabatic process under consideration.

Note that $\ln k$ can be identified with the entropy of the system in the energy eigenstate $\psi_k$ \cite{HHD}.
The sustained linear growth of entropy (hence - exponential growth of energy) in classical systems with periodically divided
configuration space was described in Refs. \cite{GRST11,GRT}. The difference with the quantum acceleration described here is that
the mechanism of the classical acceleration is the loss of the ergodicity in the phase space, which leads to the destruction of the
adiabatic invariance of the entropy. This works in a universal fashion for other non-ergodic classical systems. In the quantum case,
we have infinitely many adiabatic invariants - the populations $I_k$, $k=1,\dots, +\infty$, of the instantaneous energy levels
($I_k=|\langle \psi(t) | \psi_k(\varepsilon t)\rangle|^2$, where $\psi(t)$ is the wave-function and $\psi_k$ are the instantaneous energy eigenstates).
So, the sustained entropy and energy growth is possible only if all of these adiabatic invariants are destroyed, and
the mere ergodicity violation of the classical limit does not seem to be enough for this.

The exponential energy growth is not guaranteed in the special case $\rho=0$, and more subtle effects are possible. For an example, we consider
a non-relativistic spin-1/2 particle in a time-dependent, spatially inhomogeneous, strong magnetic field. We scale the Planck constant, the mass, and the charge
of the particle to $1$, and consider the slowly changing vector potential
${\bf A}(x,y,z,\varepsilon t)=(-y B(\varepsilon t) - F(z,\varepsilon t), x B(\varepsilon t)+G(z,\varepsilon t), 0)$
and scalar electric potential $\phi=V(z)-{\bf A}^2/2$. As ${\rm div} {\bf A}=0$, the Pauli equation takes the form
$$i\frac{\partial}{\partial t} \psi_{\pm} = \frac{1}{2}(-\triangle + 2 i {\bf A} \cdot {\bf \nabla})\psi_{\pm} + V(z)\psi_\pm-
\frac{1}{2} {\bf\sigma} \cdot {\bf B}(z,\varepsilon t) \left(\begin{array}{c}\psi_+\\\psi_-\end{array}\right),$$
where $\bf\sigma$ are the Pauli matrices and ${\bf B}=(F_z(z,\varepsilon t), G_z(z,\varepsilon t), 2B(\varepsilon t))$ is the magnetic field. We consider only the
wave-functions $\psi_\pm$ which are independent of $(x,y)$. In other words, by imposing the boundary condition $\psi=0$ at $x^2+y^2=\delta^2$,
we confine the particle to the $\delta$-thin infinite cylinder around the $z$-axis, so in the limit $\delta\to 0$ the problem becomes one-dimensional, and the equation takes
the form
\begin{equation}\label{mf}\!\!\!\!
i\frac{\partial \psi_{\pm}}{\partial t}\! = -\frac{1}{2}\frac{\partial^2\psi_{\pm}}{\partial z^2} + V(z)\psi_\pm-\!\!\left(\!\!\begin{array}{cc}
B(\varepsilon t) & C(z,\varepsilon t)\\ C^*(z,\varepsilon t) & B(\varepsilon t)\end{array}\!\!\right)\! {\bf\psi},\!
\end{equation}
where $C=(F_z-iG_z)/2$. Computations become explicit if we take $V(z)=\frac{1}{2}z^2$. Let the magnetic field oscillate periodically as a function of the slow
time $\tau=\varepsilon t$, so that for some interval $[\tau_1,\tau_2]$ during the period the $x$ and $y$ components of the magnetic field vanish. Thus, the coefficient
$C$ in Eq. (\ref{mf}) vanishes and the components $\psi_+$ and $\psi_-$ of the wave function evolve independently at $\tau\in [\tau_1,\tau_2]$. During the rest of the
oscillation period, we assume that $C$ is a non-constant function of $z$, i.e. the magnetic field direction varies with $z$, so $\psi_+$ and $\psi_-$ become coupled.
This means that the system has an additional quantum number (spin up or spin down) at $\tau\in[\tau_1,\tau_2]$, while for the rest of the period this quantum number is destroyed.

According to the theory above, the slow time evolution of such system is described by the values the energy level number $k$ takes at the beginning of each oscillation
period; the trajectories of the level numbers are completely determined by the energy spectra at the moments of separation and reconnection, $\tau_1$ and $\tau_2$.
In our case, the spectrum at $\tau=\tau_j$ is given by $E_m^+=m-1/2-B(\tau_j)$ for the ``up''-states and $E_n^-=n-1/2+B(\tau_j)$ for the ``down''-states ($m,n\geq 1$).
We take $B(\tau_1)=-1/4$ and $B(\tau_2)=3/4$. Thus, if we order the spectrum by the increase of the energy, the up and down states will alternate. Namely,
the energy levels with even number $k$ correspond to the up states and odd $k$ correspond to the down states, except for the ground state ($k=1$) at the moment of reconnection
$\tau=\tau_2$, which is also the up state. In terms of the indicator sequences $\sigma_j(k)$ from formula (\ref{kbk}) this reads as $\sigma_1(k)=(-1)^k$ and
$\sigma_2(1)=1$, $\sigma_2(k)=(-1)^k$ at $k\geq 2$, and their sums are given by $S_1(k)=-1,0,-1,0,-1,...$ and $S_2(k)=1,2,1,2,1,2,...$. Thus,
if $\sigma_1(k)=\sigma_2(\bar k)$, then either $k$ and $\bar k$ are of the same parity, so $S_1(k)-S_2(\bar k)=-2$, or $k$ is even and $\bar k=1$, which gives
$S_1(k)-S_2(\bar k)=-1$. By (\ref{kbk}), we obtain the following law for the change of the energy level number after one period:\\
spin up - $\; \bar k= k-2$ for even $k\geq 4$, $\;\bar k=1$ if $k=2$, and $\bar k=k+2$ if $k$ is odd (spin down).

As we see, in the process under consideration, the non-homogeneous magnetic field makes slow, large amplitude oscillations
in time in such manner that it becomes spatially homogeneous during a part of the oscillation period. This may lead to a linear growth of the particle energy
for the states with a certain spin orientation (down states in our example), while in the states with the opposite spin a constant amount of energy will be lost
with each period, until a minimal value of energy is reached, after which the spin orientation changes and the eternal acceleration starts.

The conclusion is that a slow periodic change of parameters of a quantum mechanical system leads, typically, to an exponential growth of energy (due to
an adiabatic level crossing) provided an additional quantum number is created and destroyed during the oscillation period. The basic example of such processes
is given by systems with periodically divided configuration space. A slower rate of the energy growth is also possible in special cases (like in the example of a spin-1/2
particle in a strong, oscillating, inhomogeneous magnetic field). For particular choices of the parameters of our process the energy growth may be preceded by
a period of cooling. This happens when for a group of energy states with a particular, ``wining'' value of the quantum number the system accelerates and,
in the adiabatic approximation, the state of the system remains in this group after each oscillation period. Then a state which does not belong to this winner
group has to lose energy  until a certain minimal value of energy is reached, after which the quantum number changes and the state joins the winner group.
Note that this leads to the desertion of energy states with certain values of the quantum number: for an arbitrary initial superposition of states
of finite energy, the system evolves with time to the superposition where most of the contribution is given by the energy states
with the winning value of the quantum number only.

I am grateful to V.Rom-Kedar, S.Zelik, A.Kirillov, U.Smilyansky, V.Gelfreich, and A.Vladimirov for very useful conversations. 
This work was supported by grant 14-41-00044 by the RSF. The support by the Royal Society grant IE141468 is also acknowledged.

\end{document}